\documentclass[conference]{IEEEtran}

\usepackage{cite}
\usepackage{graphicx} 
\PassOptionsToPackage{hidelinks}{hyperref}
\usepackage{hyperref}
\usepackage{tikz} 
\usetikzlibrary{arrows.meta, positioning, shapes.geometric, calc}
\usepackage{amsmath, amssymb, amsfonts} 
\usepackage{booktabs} 
\usepackage{textcomp}
\usepackage{xcolor}
\usepackage{float}
\usepackage{tabularx}
\usepackage{supertabular}
\usepackage{caption}
\usepackage{placeins}
\usepackage{algorithm}
\usepackage{algpseudocode}
\renewcommand{\thetable}{\arabic{table}} 
\usepackage{multirow} 
\usepackage{pgfplots} 
\usepackage{textcase}
\usepackage{makecell}
\usepackage[nameinlink,capitalise,noabbrev]{cleveref}
\crefname{table}{Table}{Tables}
\makeatletter
\def\fnum@table{\NoCaseChange{Table}~\thetable} 
\makeatother
\usepackage[compact]{titlesec}
\titlespacing{\section}{0pt}{*1}{*1} 

\pgfplotsset{compat=1.18}

\title{LLMGreenRec: LLM-Based Multi-Agent Recommender System for Sustainable E-Commerce}

\author{
    \IEEEauthorblockN{Hao N. Nguyen\IEEEauthorrefmark{1}, Hieu M. Nguyen\IEEEauthorrefmark{2}\IEEEauthorrefmark{3}, Son Van Nguyen\IEEEauthorrefmark{1},  Nguyen Thi Hanh\IEEEauthorrefmark{2}}
    \IEEEauthorblockA{
      \IEEEauthorrefmark{1}Faculty of Computer Science, PHENIKAA University, Hanoi, Vietnam; \\
      \IEEEauthorrefmark{2}Faculty of Interdisciplinary Digital Technology, PHENIKAA University, Hanoi, Vietnam; \\
       \IEEEauthorrefmark{3}Corresponding author at PHENIKAA University, Hanoi, Vietnam.} 
         \IEEEauthorblockA{
    \IEEEauthorrefmark{1}22010115@st.phenikaa-uni.edu.vn,
    \IEEEauthorrefmark{2}\IEEEauthorrefmark{3}hieu.nguyenminh1@phenikaa-uni.edu.vn, \\
    \IEEEauthorrefmark{1}son.nguyenvan@phenikaa-uni.edu.vn,
    \IEEEauthorrefmark{2}hanh.nguyenthi@phenikaa-uni.edu.vn
    }}

\begin{document}

\maketitle

\begin{abstract}
Rising environmental awareness in e-commerce necessitates recommender systems that not only guide users to sustainable products but also minimize their own digital carbon footprints. Traditional session-based systems, optimized for short-term conversions, often fail to capture nuanced user intents for eco-friendly choices, perpetuating a gap between green intentions and actions. To tackle this, we introduce LLMGreenRec \footnote{Our source code and data are available at: \url{https://github.com/haongocng/LLMsGreenRec}}, a novel multi-agent framework that leverages Large Language Models (LLMs) to promote sustainable consumption. Through collaborative analysis of user interactions and iterative prompt refinement, LLMGreenRec's specialized agents deduce green-oriented user intents and prioritize eco-friendly product recommendations. Notably, this intent-driven approach also reduces unnecessary interactions and energy consumption. Extensive experiments on benchmark datasets validate LLMGreenRec's effectiveness in recommending sustainable products, demonstrating a robust solution that fosters a responsible digital economy.
\end{abstract}


\section{Introduction}
In the context of e-commerce boom, every user click and search query not only is a consumer action, but also consumes energy resources to power data centers. When multiplied by billions of daily interactions globally, the carbon footprint of online shopping becomes a significant concern. Consequently, a dual challenge emerges: guiding users toward environmentally friendly products while simultaneously optimizing the interaction process itself to minimize energy waste.

Evidence shows a clear and growing consumer awareness of environmental issues, as well as a demand for sustainable products. According to the PwC's Voice of the Consumer Survey 2024 report, 54\% of Vietnamese consumers are willing to pay up to 10\% more for a product made from recycled or sustainable materials \cite{pwc2024voice}. Internationally, the Global 2024 Gen Z and Millennial Survey by Deloitte highlights that two in 10 Gen Zs and millennials have changed jobs to better align work with their environmental values, and both demographic cohorts are also willing to pay more for sustainable products \cite{deloitte2024genz}. Nevertheless, a significant contradiction persists: the stated sustainable consumption intentions of most customers do not align with their actual purchasing behavior. A BCG's 2022 survey conducted on 19,000 consumers across eight countries indicates that, while up to 80\% of consumers claim they think about sustainability in their daily purchases, only 1--7\% alter their purchasing habits to reflect this \cite{bcg2022green}.

This say-do gap is most evident when a well-meaning user begins an online shopping session. Their good intentions are often thwarted by the complex structure of e-commerce platforms, where they face an overwhelming volume of information and a predominance of conventional products. This makes finding and evaluating eco-friendly alternatives difficult, leading to decision fatigue, and ultimately a retreat to more familiar and convenient options. This issue not only pertains to the consumer side but also poses a technical challenge for e-commerce businesses in a green economy---every user's action scroll or search consumes server energy. While a single search or scroll seem negligible, when multiplied, they contribute drastically to the digital carbon footprint.

To optimize the search process, reduce redundant actions, and shorten the user's journey, recommender systems have been developed and deployed. Nonetheless, traditional systems exhibit inherent limitations when measured against the goals of a green economy. In particular, they are optimized to predict the most likely next action based on popularity data, inadvertently creating an uneven playing field. Providers of sustainable products, often smaller-scale enterprises with limited resources, struggle to compete for attention in a digital marketplace dominated by major brands. Existing recommender systems frequently overlook these niche products, rendering them nearly invisible to their target potential customers.

The root of this problem lies in the shortcomings of current session-based recommender systems. They are designed mainly to maximize short-term conversion rates, lacking the sophistication to recognize deeper user intents, such as a desire to make responsible choices. Therefore, there is an urgent need for a new generation of recommender systems that are not only intelligent in capturing user intents but also flexible enough to integrate environmental criteria into their suggestion process. Developing such systems is not merely a technical improvement, but a strategic and vital step to break down existing barriers, foster the market for green products, and realize a sustainable and responsible digital economy.

To address this need, this paper focuses on applying an advanced approach: building a high-performance, session-based recommender system for identifying user intents and prioritizing green products. By applying Large Language Models (LLMs) with their complex reasoning capabilities, the system can accurately predict user needs with fewer interactions, thereby shortening search times and reducing operational energy consumption. Concurrently, the system will leverage this intent understanding to proactively introduce environmental-friendly alternatives. The ultimate goal is to create a shopping process that is not only intelligent and convenient but also energy-efficient and environmentally responsible, contributing practically to the growth of a green economy.

\section{Related Work}
Intent-based session recommendation is not an entirely new topic in the field of recommender systems. From early work on session-based recommender systems, researchers have gradually shifted towards leveraging user intents to improve recommendation accuracy. With the development of deep learning and, more recently, Large Language Models, the field has seen significant advancements, particularly in integrating sustainability criteria to recommend green products and promote eco-friendly consumption. The following sections will review related research, including traditional session-based recommender systems, intent-based session recommendations, applications of LLMs, multi-agent systems, and studies on green product recommendations.
\begin{figure*}[ht]
  \centering
  \includegraphics[width=0.6\textwidth]{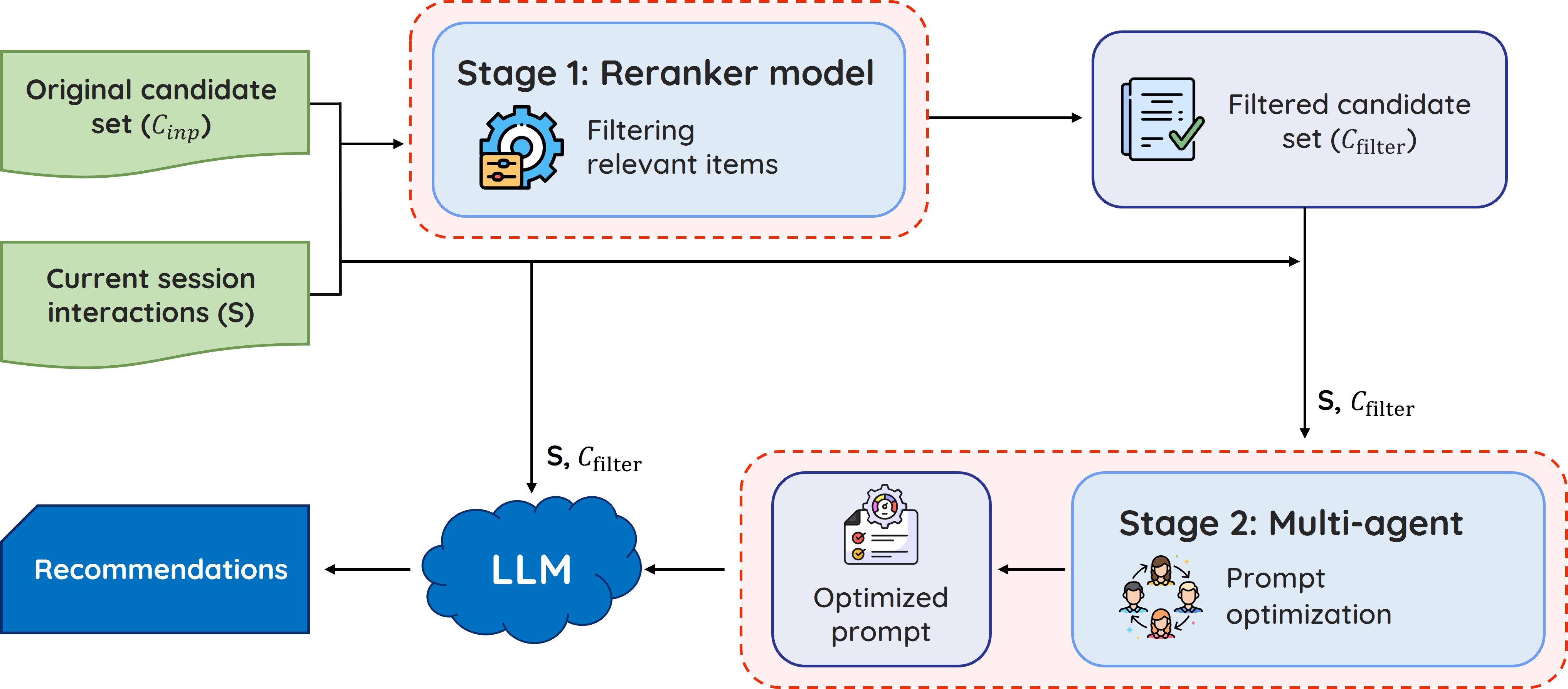}
  \caption{Overall workflow of LLMGreenRec}
  \label{fig:overall_workflow}
\end{figure*}
\subsection{Session-Based Recommender Systems}
Session-based recommender systems focus on analyzing a user's short-term sequence of actions within a specific session, typically without relying on long-term historical data or user profiles. In 1999, Schafer et al. \cite{schafer1999ecom} proposed using interaction history and personal preferences to suggest products, aiming to increase sales on e-commerce platforms such as Amazon and eBay. This group of authors \cite{schafer2001ecom} later developed new methods that exploited relationships between products and user behaviors to enhance the prediction of customer needs. 

One of the major challenges for early recommender systems was the massive volume of data. In 2000, Sarwar et al. \cite{sarwar2000dimRedRecSys} proposed a Singular Value Decomposition (SVD) algorithm to reduce data dimensionality, thereby decreasing computation time and generating potential product lists more effectively. By 2002, the same group \cite{sarwar2002largeScaleEcom} introduced a method of clustering customers, based on a nearest-neighbor algorithm combined with collaborative filtering, to improve accuracy. Huang et al. \cite{huang2004graphModelEcomRS} later innovated this approach by using multi-relational graphs, where links between products and users were represented by weights, delivering superior performance when integrated with other recommendation techniques.

The advent of deep learning marked a significant paradigm shift in session-based recommendation approaches. In 2016, Hidasi et al. \cite{hidasi2016gru4rec} were the first to apply hierarchical Recurrent Neural Networks (RNNs) to uncover latent features within short-term session data. Following this success, Tan et al. \cite{tan2016improvedRNN} optimized RNNs to process session data more rapidly, achieving noticeable results. In 2017, Li et al. \cite{li2017neuralAttentiveSBR} augmented RNNs with an attention mechanism, allowing the model to focus on the most important items within a session. Wu et al. \cite{wu2019srGNN} enhanced session-based recommender systems by introducing SR-GNN, a model that utilizes Graph Neural Networks (GNNs) to represent a session as a graph, thereby better capturing the relationships between items. In recent years, researchers have noted that RNNs are not optimal for sessions with high randomness---Gwadabe et al. \cite{gwadabe2022GRASER} proposed GRASER, a GNN-based model, to overcome this limitation.

\subsection{Intent-Based Session Recommender Systems}
Intent-based session recommendation extends the session-based problem by focusing on identifying user intent within each session, such as interacting with a product or searching for information. Methods in this area can be divided into two main categories: those assuming a single user intent and those designed to learn multiple intents.

For the single-intent assumption, STAMP \cite{liu2018STAMP} emphasized user's final action as the primary intent signal. MSGAT \cite{qiao2023MSGAT} improved upon this by incorporating both local and global information from similar sessions. However, this approach fails to consider that a single session may contain multiple intents, potentially disregarding critical information.

To address this, subsequent research has focused on identifying multiple intents within the same session. NirGNN \cite{jin2023dualIntentGNN} used an attention mechanism to learn multiple intents. The MCPRN method \cite{wang2019MCPRN} applied routing channels to classify the intent of each item. IDSR \cite{chen2020intentDiversifiedSR} projected item representations into multiple intent spaces, whereas HIDE \cite{li2022HIDE} partitioned item embeddings into segments representing different intents. Moreover, MIHSG \cite{guo2022multiGranularityIntent} and STAGE \cite{li2022STAGE} learned intents at varying levels of granularity. Despite being advanced, these approaches tend to assume a fixed number of intents per session and lack transparency in their recommendation process.

\subsection{Sustainable Recommender Systems}
Sustainable recommender systems focus on encouraging users to select eco-friendly products based on data related to environmental impact and green consumption behavior. Early research in this area primarily used collaborative filtering and machine learning techniques to classify products based on green indices such as carbon emissions and recycled content. A comprehensive survey on recommender systems for sustainability indicated that these systems not only reduce energy consumption but also support the achievement of the UN's Sustainable Development Goals (SDGs) by prioritizing products with green certifications \cite{felfernig2023sustainabilityRS}.

The integration of Artificial Intelligence (AI) has led to higher efficacy in green product recommendations. Dinesh et al. \cite{dinesh2024ecoconscious} proposed a AI-driven, eco-friendly fashion recommender system to analyze user behavior and suggest apparel made from recycled materials, helping reduce the fashion industry's environmental impact. Similarly, a hybrid recommender systems that combined product functional data with sustainability profiles of their components have been developed to suggest green alternative materials in consumer goods manufacturing \cite{chizzali2025ecoingredientCaseStudy}. Nevertheless, the lack of standardized green index data and the complexity of context remain key limitations, underscoring the need for advanced models such as LLMs to boost accuracy and personalization.

\subsection{Applications of Large Language Models (LLMs)}
The emergence of LLMs such as GPT and LLaMA has opened up new avenues for session-based and intent-based recommender systems, thanks to their superior semantic and contextual understanding. The Zero-Shot Next-Item Recommendation study \cite{wang2023zeroShotNIR} was one of the first to use zero-shot prompting to apply an LLM to a session-based recommender system without model fine-tuning. Hou et al. \cite{hou2024LLMzeroShotRankers} exploited the in-context learning capabilities of LLMs, using them as ranking tools based on the order and recency of items in a session. Moreover, models such as BIGRec \cite{bao2025BIGRec} and GPT4Rec \cite{li2023GPT4Rec} have been fine-tuned to better suit the recommendation domain and accelerate system performance. Meanwhile, RecInterpreter \cite{yang2023RecInterpreter} explored how to decode LLM representations for application in sequential recommendation tasks. The PO4ISR study by Sun et al. \cite{sun2024LLMIntentSessionRec} marked a significant step forward by using ChatGPT to improve intent-based session recommender systems through automated prompt optimization. Following PO4ISR, the MACRec technique \cite{wang2024MACRec} was proposed as a further advancement, focusing on the use of multiple collaborating agents in the recommender system. MACRec leveraged the coordination of multiple LLMs, where each can assume a different role to enhance overall performance.

Recently, LLMs have been applied to enhance sustainable recommender systems, leveraging their natural language processing capabilities to analyze users' green intentions and eco-friendly product data. One eco-friendly product recommender system used Llama-2 as its core engine, generating personalized embeddings from natural language queries and suggesting products based on green factors such as recycled materials and low carbon footprint, thereby increasing user satisfaction and promoting sustainable consumption \cite{bondgulwar2025EcoFriendlyRecsys}. Likewise, Zhou et al. have demonstrated that LLMs can reduce energy consumption via recommendations optimization, ultimately facilitating global environmental goals \cite{zhou2024AdvSustRecSurvey}. In a more specific domain, a recommender system for prosumers used Scikit-llm and zero-shot classifiers to evaluate sustainability scenarios, such as recommending renewable energy products based on real-time data \cite{oprea2024RecSysProsumers}. Furthermore, LLMs have been integrated into sustainable tourism systems, using Retrieval-Augmented Generation (RAG) to enhance contextual relevance and reduce environmental impact \cite{banerjee2025EnhancingTourismRec}. These studies highlight the potential of LLMs in processing complex sustainability data.

\section{Methodology}
This section introduces a novel architecture, LLMGreenRec, designed to recommend high-quality and relevant green products. As shown in Figure \ref{fig:overall_workflow}, there are two key inputs: the original candidate set ($C_{\text{inp}}$) and current session interactions ($S$). These inputs are processed through two stages: \textbf{Reranker model} and \textbf{Multi-agent}. The first stage reduces the size of $C_{\text{inp}}$ by filtering out irrelevant items based on session interactions, yielding a more focused candidate set ($C_\text{filter}$). The second stage employs a multi-agent system, where specialized agents assess user relevance and sustainability attributes separately. A final LLM then reasons over these outputs to select the optimal set of products that balances both criteria.
\subsection{Filtering Relevant Items with Reranker Model}
\FloatBarrier 
\begin{figure}[!h]
  \centering
  \includegraphics[width=0.9\linewidth]{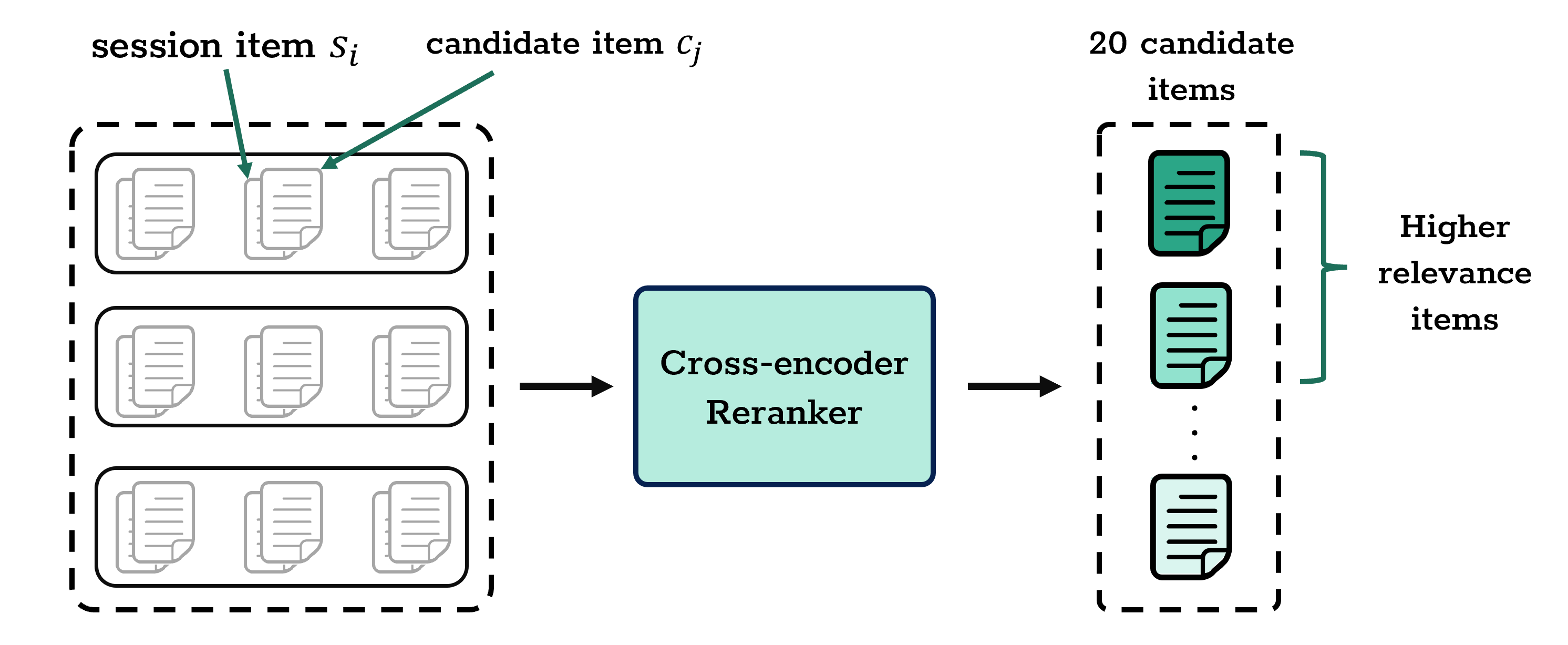}
  \caption{Overall pipeline of Cross-encoder reranker}
  \label{fig:stage1_workflow}
\end{figure}
\FloatBarrier
The first stage of the overall architecture is a Cross-encoder reranker \cite{reimers2019sentencebert}, which serves as a relevance filtering step, as illustrated in Figure \ref{fig:stage1_workflow}. This model processes two key inputs: the chronological current session interaction, $S = [S_1, S_2,\dots, S_n]$, where each $s_i$ represents an item interacted by the user and the initial candidate set, $C_\text{inp} = [c_1, c_2,\dots, c_\text{100}]$., is formed by randomly selecting 100 products from the entire item catalog, ensuring the target item is always included. The Cross-encoder operates by systematically evaluating the relevance of each candidate item in $C_\text{inp}$ with respect to the user's session interactions $S$. First, the filtering process begins with extracting items from the current session $S$ and the initial candidate set $C_\text{inp}$. Afterward, for each item \(s \in S\), the Cross-encoder pairs it with each candidate item \(c \in \mathcal{C}_{\text{inp}}\) to form a pair \((s, c)\). A pre-trained Cross-encoder model, consisting of a Sentence Transformer and a Classifier, processes each pair. The Sentence Transformer generates a joint embedding that captures the semantic relationship between \(s\) and \(c\), and the Classifier outputs a relevance score, $\mathrm{score}_{\text{pair}}(s, c)$, ranging from 0 to 1. For each candidate item \(c\), the overall relevance score is computed as the average of the pairwise scores across all session items:
    \[
        \mathrm{score}(c) \;=\; \frac{1}{\lvert S \rvert} \sum_{s \in S} \mathrm{score}_{\text{pair}}(s, c).
    \]
This averaging approach ensures that the relevance of a candidate item reflects its alignment with the entire session context. Finally, the candidate items in $C_\text{inp}$ are ranked based on their average relevance scores, and the top 20 items with the highest scores are selected to form the filtered candidate set \(\mathcal{C}_{\text{filter}}\). This step reduces the candidate pool while preserving the most pertinent items. The output of this stage is a filtered candidate set, $C_\text{filter}=[c_1',c_2',\dots,c_\text{20}']$, containing the top 20 items deemed most relevant to the session $S$. This reduction from 100 to 20 candidates ensures that subsequent stages can focus computational resources on a high-quality subset.

\subsection{Prompt Optimization with Multi-Agent Framework}
\begin{figure*}[ht]
  \centering
  \includegraphics[width=0.6\textwidth]{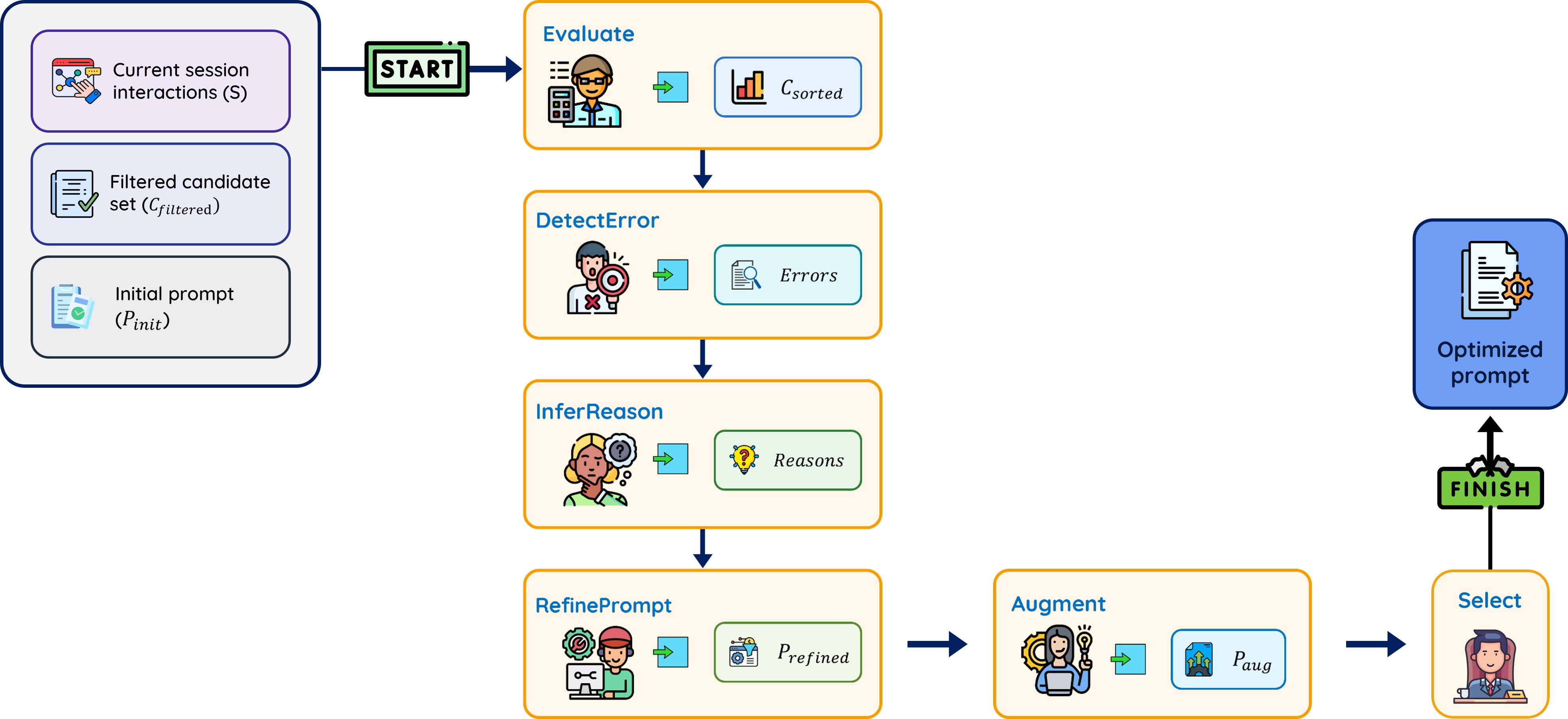}
  \caption{Overall pipeline of the multi-agent system}
  \label{fig:stage2_workflow}
\end{figure*}
To address the dual challenge of meeting user needs while promoting green consumption, the second stage is a multi-agent Large Language Model (LLM) framework, as displayed in Figure \ref{fig:overall_workflow}. This system is designed to understand and predict user's true intent, optimizing the recommendation process with a special emphasis on green products. The core of the system is an automated prompt optimization process that leverages the superior reasoning capacity of LLMs. The ultimate goal is to create a recommender system that is not only highly accurate but also capable of proactively and naturally suggesting sustainable alternatives, helping bridge the gap between green consumption intention and actual user behavior.

For input, the system takes product information with which a user has interacted during the current session, along with a candidate list of 20 products that have been filtered from stage 1. This refined set of candidates serves as the starting point, allowing the LLM agents to focus on the most relevant options instead of a large, unoptimized dataset.

Based on this reduced candidate set, a pipeline orchestrating six LLM-based agents is designed to analyze the semantics of session interactions and predict the next items of interest to the user. As illustrated in Figure \ref{fig:stage2_workflow}, the six agents are \textbf{Evaluate, DetectError, InferReason, RefinePrompt, Augment} and \textbf{Select}. Each agent has a specific function and task, and they work in close coordination, continuously improving the prompt based on feedback from the prediction process to enhance the system's overall performance.

\subsubsection{Evaluate Agent}
Evaluate Agent's inputs include the initial prompt, session data, and the refined candidate list. Its output is a ranked list of products, typically represented as a text string or JSON object for easy processing in subsequent steps. This agent serves as a crucial initial test, allowing the system to assess and adjust the initial prompt for personalized optimization. Evaluate Agent harnesses LLM's advanced natural language processing and semantic reasoning, enabling it to understand complex contexts from user interaction history. For instance, if a user views products related to a specific domain, the LLM can infer their intent and prioritize similar items. Compared to traditional methods (e.g. collaborative filtering), LLMs offer greater flexibility and more accurate predictions, especially with sparse, long-term historical data. Furthermore, the agent uses session data to personalize the ranked list for each user, focusing on recent behavior rather than general trends and increasing the likelihood of relevance. The ranked list produced by this agent is a critical input for other agents, such as DetectError and RefinePrompt,  creating a seamless prompt optimization workflow. Evaluate Agent is part of an iterative system where feedback from subsequent agents is used to improve the prompt in a positive feedback loop, making the system progressively more accurate over time.

\subsubsection{DetectError Agent}
The ranked list from Evaluate Agent is then passed to DetectError Agent for error checking. If the ground-truth target item is ranked below a certain threshold (e.g., outside the top 10 of the candidate list), the agent flags it as an error case. The agent's inputs are the predicted ranked list and the actual position of the target item; its output is a simple flag indicating whether the session is an error or not. The strength of this agent lies in its sensitivity to deviations, letting the system react quickly to ineffective prompts and maintain a high quality of recommendations. In an LLM-based system, the prompt is central to how the model interprets data, but it may not always be optimal. Without a mechanism to detect these failures, the system could continue using suboptimal prompts, leading to inaccurate suggestions and a degraded user experience. This agent acts as a quality control filter, swiftly identifying when a prompt is underperforming and signaling other agents to make timely adjustments.

\subsubsection{InferReason Agent}
After an error is detected, InferReason Agent analyzes the root cause. Its inputs are the error report from DetectError Agent, the current prompt, and the interaction history. It then deduces potential reasons, such as ``The prompt was not context-specific enough" or ``The prompt overlooked recent user behaviors". The agent's output is a list of logical hypotheses explaining the error, generated via the LLM's self-reflection mechanism. For instance, upon receiving an error, the agent might reason: ``Why was the target item not in the top 10?", then use the LLM to analyze the input data's semantics to conclude: ``The prompt did not emphasize the user's preference for high-tech products". This reasoning helps the system not only to correct errors but also to learn from its mistakes, providing a well-founded basis for prompt improvement. The agent provides detailed inferences that help the system adjust the prompt effectively, allowing the system to become more intelligent and adaptive without continuous manual intervention. In the long run, this agent optimizes the prompt refinement process through targeted and logical hypotheses, saving runtime and improving ranking quality compared to random trial-and-error.

\subsubsection{RefinePrompt Agent}
Based on the reasons derived from InferReason Agent, RefinePrompt Agent is tasked with modifying the prompt to address the identified issues. Its inputs are the old prompt and the list of reasons; its output is a new, improved prompt. The agent adjusts wording or adds contextual information, such as the product category or session focus. The improved prompt is encapsulated within \texttt{<START>} and \texttt{<END>} tags for easy extraction. The agent's flexibility facilitates creation of more specific prompts that correctly reflect the session's intent and context, thereby improving ranking performance in subsequent optimization loops. This agent is crucial, as it actively improves the prompt based on well-founded analyses, ensuring that adjustments target the root cause of the problem. This makes the prompt more effective and adaptable to specific situations, moving beyond a one-size-fits-all approach.

\subsubsection{Augment Agent}
To enhance the exploration of prompt variations, Augment Agent generates variants of the refined prompt. Taking the prompt from RefinePrompt Agent as input, this agent produces 3--5 different versions that preserve the core meaning but alter the phrasing. For example, if the original prompt was ``Sort products by technology interest from interaction history," Augment Agent might generate ``Rank products based on recent tech-related shopping intent". This is achieved by leveraging the paraphrasing capability of LLM. This agent introduces linguistic diversity and expands the search space for optimal phrasing for specific use cases. By proactively expanding the search space, Augment Agent prevents the system from becoming stuck with a single mode of expression, allowing it to adapt to diverse and complex scenarios.

\subsubsection{Select Agent}
After Augment Agent creates prompt variations, Select Agent chooses the best one to use. It employs the UCB (upper confidence bound) \cite{pryzant2023APO} algorithm to balance two objectives:
\begin{itemize}
    \item{Exploitation:} Choose prompts that already perform well.
    \item{Exploration:} Test new prompts to see if they are better.
\end{itemize}
The agent calculates a UCB value for each prompt as follows:
\[
\text{UCB} = \frac{\text{reward}}{\text{selection}} 
+ \text{explore\_param.} \, \sqrt{\frac{\log(t)}{\text{selection}}}
\]
Reward represents the prompt's quality by assessing whether the user's items of interest are ranked highly, selection is the number of times that prompt has been tested, t is the total number of trials conducted for all prompts, and explore\_param adjusts the level of exploration for new prompts.

The prompt with the highest UCB value is selected for the next iteration, as detailed at Algorithm \ref{alg:ucb}. This method prevents the system from getting stuck with a single prompt and encourages continuous improvement by exploring new options, thereby enhancing its predictive capabilities over time.
\begin{algorithm}
\caption{UCB}
\begin{algorithmic}
\State $P = \{p_1, p_2, \ldots\}$, $R[p] = 0$, $S[p] = 0$ for all $p \in P$, $T = 0$
\For{$i = 1$ to max\_trials}
    \State $T \leftarrow T + 1$
    \For{each $p \in P$}
        \State $\text{UCB}[p] = \begin{cases} 
        +\infty & \text{if } S[p] = 0 \\
        \frac{R[p]}{S[p]} + \sqrt{\frac{2\ln(T)}{S[p]}} & \text{otherwise}
        \end{cases}$
    \EndFor
    \State $\text{sel} = \arg\max_p \text{UCB}[p]$
    \State $R[\text{sel}] \leftarrow R[\text{sel}] + \text{evaluate}(\text{sel})$
    \State $S[\text{sel}] \leftarrow S[\text{sel}] + 1$
\EndFor
\label{alg:ucb}
\end{algorithmic}
\end{algorithm}

\section{Experiments and Results}
Two primary research questions are addressed:
\begin{itemize}
    \item \textbf{RQ1:} Can LLMGreenRec's multi-agent module outperform baseline models in general recommendation tasks?
    \item \textbf{RQ2:} How effective is LLMGreenRec at recommending relevant and sustainable products to users?
\end{itemize}

\subsection{Experimental Setup}
\subsubsection{Datasets}
LLMGreenRec's performance is evaluated on three real-world datasets. The first, MovieLen-1M (ML-1M)\cite{Harper2015MovieLens}, provides user ratings for movies. The other two stem from Amazon: the Games\cite{Ni2019Justifying} dataset, which contains video game ratings, and the Bundle dataset\cite{Zhu2022BundleDataset}, which offers session-based data for Electronic, Clothing, and Food categories with explicit intent annotations. Each instance in our session-based datasets is structured around two components: the current session interactions, a chronological sequence of products the user engaged with, and the target item, which is the final product the user selected. We randomly sample 100 items for each session to form the candidate set, which serves as the initial pool for our filtering and prompt optimization stages.

\subsubsection{Baselines}
LLMGreenRec is benchmarked against a comprehensive set of baselines, classified into three main types. The first category consists of conventional methods such as Mostpop \cite{ji2020popularityBaseline} for recommending popular items, SKNN \cite{diet2017SKNN} for session-level similarity, and FPMC \cite{stef2010FPMC}, a Markov chain-based matrix factorization method. The second type comprises deep learning-based methods, which are divided into single-intent models such as the RNN-based NARM \cite{li2017neuralAttentiveSBR}, the last-item-focused STAMP \cite{liu2018STAMP}, and the graph-based GCE-GNN \cite{ziyang2023GCE-GNN}; and multi-intent models such as MCPRN \cite{wang2019MCPRN}, HIDE \cite{li2022HIDE}, and Atten-Mixer \cite{pei2022Atten-mixer}, which are designed to capture more diverse user purposes. The final category consists of other LLM-based methods, including the zero-shot prompter NIR \cite{wang2023zeroShotNIR} and the recent high-performing PO4ISR \cite{sun2024LLMIntentSessionRec} model.

\subsubsection{Evaluation Metrics}
Following state-of-the-art practices in intent-driven session recommendation \cite{sun2024LLMIntentSessionRec}, two evaluation metrics are used: Hit Rate (HR@K) and Normalized Discounted Cumulative Gain (NDCG@K), with K set to 1 or 5 to assess the top-ranked and top-five items, respectively. For both metrics, higher values indicate desirable ranking results. To ensure a fair comparison, all LLM-based methods are implemented on top of the same foundational model, Meta-Llama-3 \cite{llama32024Herd}.

\subsection{Experimental Results}
\subsubsection{Overall Comparison (RQ1)}
\begin{table*}[t]
  \centering
  \captionsetup{font=small}
  \caption{Comparison of recommendation methods on three datasets, with the best results highlighted in bold and `--' for very small values}
  \label{tab:overall_comparison}
  \setlength{\tabcolsep}{2pt}
  \renewcommand{\arraystretch}{1.15}
  \resizebox{1\textwidth}{!}{%
  \begin{tabular}{l l ccc ccc ccc ccc}
    \toprule
    \multirow{2}{*}{\textbf{Datasets}} & \multirow{2}{*}{\textbf{Metrics}}
      & \multicolumn{3}{c}{\textbf{Traditional Methods}}
      & \multicolumn{3}{c}{\textbf{Single-Intent Methods}}
      & \multicolumn{3}{c}{\textbf{Multi-Intent Methods}}
      & \multicolumn{3}{c}{\textbf{LLM-based Methods}} \\
    \cmidrule(lr){3-5}\cmidrule(lr){6-8}\cmidrule(lr){9-11}\cmidrule(lr){12-14}
    & & MostPop & SKNN & FPMC
      & NARM & STAMP & GCE-GNN
      & MCPRN & HIDE & Atten-Mixer
      & NIR & PO4ISR & \textbf{LLMGreenRec} \\
    \midrule
    \multirow{4}{*}{ML-1M}
      & HR@1    & 0.0004 & 0.1270 & 0.1132 & 0.1692 & 0.1584 & 0.1312 & 0.1434 & 0.1498 & 0.1490 & 0.0572 & 0.2000 & \textbf{0.2550} \\
      & HR@5    & 0.0070 & 0.3600 & 0.3748 & 0.5230 & 0.5078 & 0.4748 & 0.4788 & 0.4998 & 0.4932 & 0.2326 & 0.5510 & \textbf{0.6110} \\
      & NDCG@1  & 0.0004 & 0.1270 & 0.1132 & 0.1692 & 0.1584 & 0.1312 & 0.1434 & 0.1498 & 0.1490 & 0.0572 & 0.2000 & \textbf{0.2550} \\
      & NDCG@5  & 0.0053 & 0.2530 & 0.2464 & 0.3501 & 0.3367 & 0.3044 & 0.3157 & 0.3256 & 0.3216 & 0.1436 & 0.3810 & \textbf{0.4391} \\
    \midrule
    \multirow{4}{*}{Games}
      & HR@1    & --     & 0.0020 & 0.0498 & 0.0572 & 0.0556 & 0.0692 & 0.0522 & 0.0696 & 0.0530 & 0.1168 & 0.2588 & \textbf{0.4580} \\
      & HR@5    & --     & 0.0020 & 0.2564 & 0.2574 & 0.2586 & 0.2744 & 0.2416 & 0.2694 & 0.2472 & 0.3406 & 0.5866 & \textbf{0.7430} \\
      & NDCG@1  & --     & 0.0020 & 0.0498 & 0.0572 & 0.0556 & 0.0692 & 0.0522 & 0.0696 & 0.0530 & 0.1168 & 0.2588 & \textbf{0.4580} \\
      & NDCG@5  & --     & 0.0020 & 0.1508 & 0.1534 & 0.1555 & 0.1701 & 0.1432 & 0.1662 & 0.1475 & 0.2310 & 0.4313 & \textbf{0.6143} \\
    \midrule
    \multirow{4}{*}{Bundle}
      & HR@1    & --     & --     & 0.0398 & 0.0322 & 0.0365 & 0.0360 & 0.0360 & 0.0458 & 0.0525 & 0.0975 & 0.1697 & \textbf{0.2815} \\
      & HR@5    & 0.0042 & --     & 0.2475 & 0.2322 & 0.2352 & 0.2237 & 0.2352 & 0.2585 & 0.2644 & 0.2832 & 0.4328 & \textbf{0.5588} \\
      & NDCG@1  & --     & --     & 0.0398 & 0.0322 & 0.0365 & 0.0360 & 0.0360 & 0.0458 & 0.0525 & 0.0975 & 0.1697 & \textbf{0.2815} \\
      & NDCG@5  & 0.0021 & --     & 0.1395 & 0.1303 & 0.1339 & 0.1267 & 0.1490 & 0.1495 & 0.1549 & 0.1939 & 0.3040 & \textbf{0.4279} \\
    \bottomrule
  \end{tabular}}
\end{table*}

To address RQ1, \cref{tab:overall_comparison} presents performance comparison for general-purpose recommendation tasks. This evaluation isolates the multi-agent module of LLMGreenRec, tasking it with recommendation from a random 20-item candidate set without prioritizing sustainability. The experimental results lead to several key findings. First, the proposed LLMGreenRec module consistently and significantly outperforms all other baselines across all datasets and evaluation metrics. This superiority is particularly evident even when compared to other high-performing LLM-based methods such as PO4ISR and NIR. The substantial performance gap validates the effectiveness of the proposed multi-agent optimization design, highlighted by relative improvements over the strong PO4ISR baseline of up to 26.6\% in HR@5 (Games dataset) and 40.7\% in NDCG@5 (Bundle dataset). This underscores the necessity of capturing complex, dynamic user intents, a task at which the LLMGreenRec module excels.

\subsubsection{Sustainable Recommendation Performance (RQ2)}
\begin{table}[t]
  \centering
  \captionsetup{font=small}
  \caption{Sustainable recommendation performance on \textit{Bundle}}
  \label{tab:rq2_bundle}
  \setlength{\tabcolsep}{6pt}
  \renewcommand{\arraystretch}{1.15}
  \resizebox{0.95\linewidth}{!}{%
  \begin{tabular}{|l|c|c|c|c|}
    \Xhline{1.1pt}
    \textbf{Dataset} & \multicolumn{4}{c|}{\textbf{Metrics}} \\ \cline{2-5}
                     & \textit{HR@1} & \textit{HR@5} & \textit{NDCG@1} & \textit{NDCG@5} \\
    \Xhline{0.8pt}
    Bundle           & 0.3950 & 0.5504 & 0.3950 & 0.4715 \\
    \Xhline{1.1pt}
  \end{tabular}}
\end{table}
To answer RQ2, \cref{tab:rq2_bundle} considers the complete, two-stage LLMGreenRec architecture, assessing its effectiveness at introducing pertinent, sustainable alternatives into a user's session. Examined on the Bundle dataset, LLMGreenRec achieves strong performance, with an HR@1 of 0.3950 and an HR@5 of 0.5504. These scores indicate that the correct sustainable product is successfully ranked first in nearly 40\% of cases and appears within the top five recommendations over 55\% of the time, underscoring the model's profound efficacy. These findings offer a significant contribution to the dual challenge of sustainable e-commerce: promoting green products while minimizing the platform's energy footprint. By accurately capturing user intent, the system not only excels at introducing relevant sustainable alternatives but also streamlines the user's discovery process. This leads to a shorter, more efficient shopping journey with fewer interactions, directly reducing the associated data center energy consumption and addressing the carbon footprint of online retail.

\section{Conclusion}
Inspired by the advanced reasoning capabilities of Large Language Models, this study proposes the two-stage LLMGreenRec framework for sustainable product recommendation within user sessions. This begins with a candidate filtering phase, where a Reranker model filters and ranks potential items to create a smaller, more relevant set. The core of our framework is a collaborative multi-agent system where six specialized agents work in a closed-loop feedback cycle. This system is uniquely designed to automatically identify failures in sustainability recommendations and infer the underlying reasons, then iteratively optimize prompts to select the most effective one for the LLM. This self-correcting mechanism enables a more precise and adaptive capture of user behavior and intent. Extensive experiments on benchmark datasets demonstrate advantage of LLMGreenRec in providing high-quality, sustainable recommendations compared to existing approaches. We believe this research offers a new paradigm for value-driven e-commerce, empowering users to make more informed and sustainable choices and paving the way for more sophisticated, ethically-aligned personalization systems.

\bibliographystyle{IEEEtran}
\bibliography{citation}

\end{document}